\documentclass[draft]{agujournal2019}
\usepackage{url} 
\usepackage[inline]{trackchanges} 
\usepackage{soul}
\usepackage{amsmath,amssymb}
\usepackage{subfig}

\draftfalse
\journalname{Space Weather}

\begin{document}

\title{A Machine-learning-Based Global Thermospheric Density Forecasting Model}

\authors{Ruochen Wang \affil{1}, Xiaoli Bai \affil{1}}

\affiliation{1}{Department of Mechanical and Aerospace Engineering, Rutgers, The State University of New Jersey, Piscataway, NJ 08854, USA}

\correspondingauthor{Ruochen Wang}{ruo.chen.wang@rutgers.edu}

\begin{keypoints}
\item AETHER-P$^3$ provides thermospheric density forecasts with quantified uncertainty.
\item Forecast skill and uncertainty reliability are assessed for quiet, moderate, and extreme geomagnetic activity.
\item AETHER-P$^3$ outperforms baseline models in forecast accuracy and uncertainty reliability.
\end{keypoints}

\begin{abstract}
Thermospheric mass density governs aerodynamic drag in low Earth orbit and is a primary source of uncertainty in orbit prediction and conjunction assessment, particularly during geomagnetic disturbances. We present AETHER-P$^3$ (Accelerometer-driven Estimation of THERmospheric density – A Physics-Informed Probabilistic Prediction Platform), a machine-learning-based global thermospheric density forecasting model that provides multi-step forecasts up to 6 hours ahead using a 3-hour input window, with predictive uncertainty estimates. AETHER-P$^3$ formulates thermospheric density forecasting as a sequence-to-sequence regression task conditioned on recent space weather evolution and a user-specified sequence of future times and locations. To enhance physical consistency and generalization, AETHER-P$^3$ incorporates JB2008 and NRLMSISE-00 density estimates evaluated at future locations, along with solar, geomagnetic, and solar-wind drivers. The network employs dual recurrent encoders and an evidential Normal-Gamma output head to jointly estimate forecast mean and uncertainty. The model is evaluated using independent satellite test cases spanning quiet, moderate, and extreme geomagnetic conditions. During quiet periods, AETHER-P$^3$ achieves high forecast skill ($R>0.95$). Under moderate activity, strong skill is retained ($R\approx0.93$), with reduced physical-domain errors than empirical baseline models. During extreme storm conditions, deterministic forecast skill degrades as expected yet remains robust ($R=0.89$–$0.90$). Predictive uncertainty remains well calibrated across all regimes. These results establish AETHER-P$^3$ as a practical, low-latency, uncertainty-aware capability for thermospheric density forecasting that supports orbit prediction, drag-risk assessment, and operational decision-making over its validated altitude range of approximately 300–520 km, with highest confidence in the data-rich 400–520 km region.
\end{abstract}

\section*{Plain Language Summary}

Satellites in low Earth orbit experience air drag from the upper atmosphere. This drag depends on thermospheric density, which can change rapidly during space weather events. If density is underestimated during geomagnetic storms, satellites can lose altitude faster than expected, and orbit predictions used for collision avoidance and maneuver planning will become inaccurate.

We developed AETHER-P$^3$, a data-driven forecasting model that predicts thermospheric density up to 6 hours in advance and provides an estimate of forecast uncertainty. The model uses recent observations of the atmosphere and space weather, together with the future times and locations where forecasts are requested, enabling predictions along satellite orbits. We tested the model during quiet periods, moderate activity, and an extreme geomagnetic storm. In all cases, the model produced accurate forecasts and reliable uncertainty estimates. Compared with commonly used physics-based space weather models, AETHER-P$^3$ showed substantially smaller errors along satellite tracks during storm-time conditions. This capability can improve drag prediction and support safer and more informed satellite operations during periods of changing space weather.

\section{Introduction}

Thermospheric mass density is a fundamental parameter governing aerodynamic drag on low Earth orbit (LEO) satellites and plays a critical role in orbit determination, conjunction assessment, and mission lifetime estimation \cite{marcos1998precision}. Density variability is primarily driven by solar irradiance and geomagnetic energy input and exhibits strong spatial and temporal variability across a wide range of timescales. During geomagnetic disturbances, rapid and nonlinear density enhancements can occur, leading to large along-track orbit errors when drag is inaccurately modeled and adversely affecting spacecraft operations. For example, on February 3, 2022, a geomagnetic storm enhanced thermospheric density by 50--125\% at altitudes of 200--400 km and resulted in the loss of 38 out of 49 SpaceX Starlink satellites \cite{fang2022space}. As satellite populations in LEO continue to grow, accurate thermospheric density prediction together with reliable uncertainty quantification has become increasingly important for space situational awareness and operational decision-making.

A variety of thermospheric density models have been developed to provide density nowcasts, defined as estimates based on contemporaneous or near-contemporaneous space weather conditions. Empirical models such as NRLMSISE-00 (Naval Research Laboratory Mass Spectrometer and Incoherent Scatter Radar model) and JB2008 (Jacchia–Bowman 2008 thermospheric density model) parameterize density using statistical relationships derived from historical observations and solar and geomagnetic indices \cite{picone2002nrlmsise, bowman2008new}. These models are computationally efficient and robust, making them widely used in operational settings; however, their reliance on simplified driver representations and climatological fits can degrade performance during periods of strong or rapidly evolving geomagnetic activity. Physics-based general circulation models simulate the coupled thermosphere--ionosphere system using first-principles equations and are capable of representing storm-time dynamics more explicitly \cite{burns1995geomagnetic, guillas2009bayesian, burns2004solar}. In practice, their predictive skill is limited by uncertainties in external forcing, model-form errors, and computational cost. Data-assimilation frameworks combine physical models with observations to improve specification accuracy, but their performance depends on observation availability and latency and they are typically optimized for near-real-time specification rather than forward-looking prediction \cite{storz2005high, mehta2020real, gondelach2021real, sutton2021toward, mutschler2023physics}.

More recently, machine-learning approaches have been introduced to model thermospheric density by learning nonlinear relationships between satellite observations, orbital parameters, and space weather drivers \cite{perez2015neural, weng2020machine, licata2022machine, li2023improving}. These models can achieve high accuracy and low-latency inference once trained and provide a flexible framework for integrating heterogeneous data sources. Nevertheless, uncertainty estimates are often absent or insufficiently calibrated, limiting their applicability in risk-sensitive operational contexts. Several learning-based global prediction methods have been proposed in recent studies to obtain thermospheric density predictions with uncertainty estimates \cite{wang2024global, gao2020calibration, licata2022machine}. However, most existing machine-learning models are designed for nowcasting and require real-time driver measurements as inputs. As a result, they do not naturally support multi-step forecasting at future times and locations.

While nowcasting models are effective for reconstructing the instantaneous thermospheric state, many space operations require true forecasting capability. Collision avoidance screening, maneuver planning, and drag-sensitive mission operations rely on density predictions made several hours in advance at user-specified orbital locations. To address this need, several forecasting-oriented approaches have been developed. Data-assimilation forecasting models, such as the C/DA-NRLMSISE-00 (Calibration and Data Assimilation-NRLMSISE-00) framework, combine thermospheric observations with an empirical background model to improve the accuracy of the initial density state prior to forward prediction \cite{forootan2022forecasting}. By reducing specification error through assimilation, these models can enhance forecast performance; however, their predictive skill remains constrained by the background model used for assimilation. 

Physics-based forecasting systems, including WAM-IPE (The coupled Whole Atmosphere Model-Ionosphere Plasmasphere Electrodynamics), propagate the coupled thermosphere--ionosphere system forward in time using first-principles dynamics driven by externally specified solar and geomagnetic inputs \cite{zhan2024quantifying}. These models are capable of capturing large-scale storm-time responses and global circulation features, making them valuable for space weather situational awareness. In practice, forecast accuracy is sensitive to uncertainties in future forcing and model parameterization. Additionally, the high computational cost limits WAM-IPE updates to approximately every six hours, restricting its applicability in operational settings.

Recently, machine-learning-based forecasting models, such as the BGMA (Bidirectional Gated Recurrent Unit with Multi-Head Attention mechanism) framework, have extended data-driven approaches from density nowcasting to multi-step forecasting by learning temporal dependencies in historical density measurements and space weather drivers \cite{pan2025interpretable}. These models can achieve strong short-horizon predictive skill with low computational cost; however, they primarily provide deterministic density predictions and do not quantify forecast uncertainty, limiting their use in operational risk assessment.

This paper makes the following contributions. The primary novelty of this work lies in the development of a conditional thermospheric density forecasting framework that enables multi-step predictions at user-specified future times and locations. Building upon prior studies, the framework integrates physics-informed input design and extends evidential deep learning to the forecasting setting, enabling both improved predictive performance and calibrated uncertainty quantification. The specific contributions are as follows. First, we propose AETHER-P$^3$ (Accelerometer-driven Estimation of THERmospheric density—A Physics-Informed Probabilistic Prediction Platform), a machine-learning-based global thermospheric density forecasting framework explicitly designed for prediction with uncertainty quantification at user-specified future times and locations. The forecasting task is cast as a sequence-to-sequence regression problem in which the predicted density sequence is jointly conditioned on recent thermospheric and space-weather evolution and on a requested future time--location sequence. Second, we incorporate physically informed baselines by evaluating JB2008 and NRLMSISE-00 density estimates at the requested future locations using historical driver information, which improves robustness and generalization across satellites and orbital regimes. Third, we adopt an evidential deep learning formulation with a Normal-Gamma output to estimate both the forecast mean and predictive uncertainty in a probabilistically consistent manner, enabling calibrated uncertainty quantification suitable for risk-aware applications \cite{amini2020deep}. Finally, we validate the proposed framework using independent satellite test cases spanning geomagnetically quiet, moderate, and extreme storm conditions, demonstrating improved multi-hour forecasting skill relative to existing forecasting-oriented models while maintaining well-calibrated uncertainty estimates, thereby supporting operational thermospheric density forecasting and drag-risk assessment.

The paper is organized as follows. Section~2 describes the methodology, including the proposed AETHER-P$^3$ forecasting input framework, the network architecture, the data, and the evaluation metrics. Section~3 presents the forecasting results and discusses model performance under geomagnetically quiet, moderate, and extreme storm conditions, with comparisons to existing models. Finally, Section~4 summarizes the main findings and outlines future directions.

\section{Methodology}

This section describes the formulation, input design, and implementation of AETHER-P$^3$ for global multi-step thermospheric density forecasting with uncertainty quantification. We first cast the problem as a sequence-to-sequence regression task in which a forecast density sequence over a fixed horizon is predicted from (i) recent thermospheric and space-weather history and (ii) a user-specified sequence of future times and locations. We then introduce the proposed input framework that integrates space-weather drivers with physically informed empirical-baseline density estimates evaluated at the requested future locations. Next, we present the dual-encoder recurrent neural network architecture and the evidential Normal-Gamma output head used to jointly estimate forecast mean and predictive uncertainty. Finally, we summarize the datasets used for training and independent testing and define the accuracy and uncertainty metrics employed for evaluation.

\subsection{Input framework}

The proposed forecasting input framework is motivated by a previously developed machine-learning-based thermospheric density nowcasting model introduced by Wang and Bai \cite{wang2024global}. In that work, strong predictive performance demonstrated the effectiveness of the nowcasting input design for capturing the relationship between thermospheric density and space weather drivers. Building on these results, and accounting for the availability and temporal characteristics of input features required for true multi-step forecasting, the present study extends and adapts the original nowcasting framework to support forward prediction at user-specified future times and locations.

\begin{figure}[!t]
    \centering
    \includegraphics[width=1\linewidth]{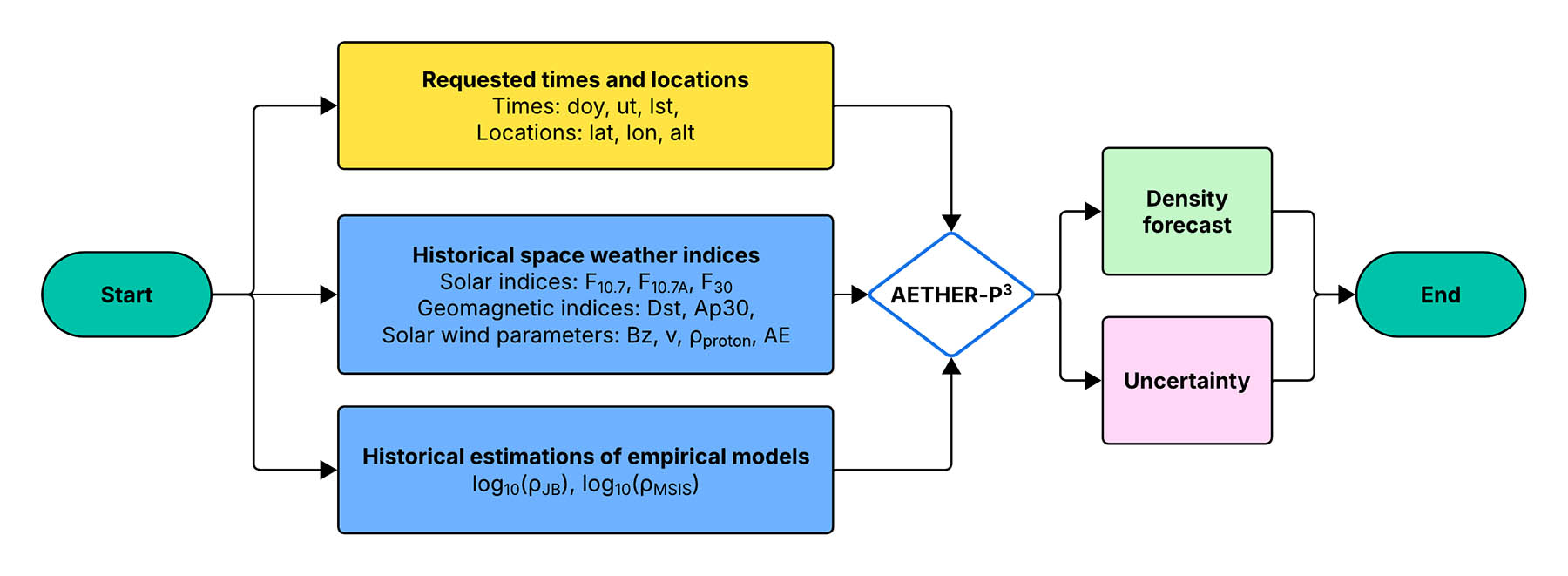}
    \caption{Global thermospheric density forecasting framework}
    \label{fig:flowchart}
\end{figure}

The thermospheric density forecasting model is formulated as a multi-step regression problem. 
Let $\hat{\boldsymbol{\rho}}_{t+1:t+n}$ denote the predicted logarithmic density sequence over a forecast horizon of length $n$. The forecasting model is expressed as
\begin{equation}
\log_{10}\!\big(\hat{\boldsymbol{\rho}}_{t+1:t+n}\big)
= f_\theta\!\left(\mathbf{X}^{(h)}_t,\mathbf{X}^{(f)}_t\right)
\end{equation}
where $\mathbf{X}^{(h)}_t$ represents the historical input sequence of length $m$, 
and $\mathbf{X}^{(f)}_t$ represents the requested future times and locations over the forecast horizon. Figure~\ref{fig:flowchart} illustrates the overall global thermospheric density forecasting framework, in which space weather drivers and empirical model baselines are integrated with requested times and locations to produce density forecasts with quantified uncertainty.

\subsubsection{Requested times and locations}
The future input sequence is defined as
\begin{equation}
\mathbf{X}^{(f)}_t =
\left[
\begin{array}{c}
\mathbf{p}_{t+1}\\
\mathbf{p}_{t+2}\\
\vdots\\
\mathbf{p}_{t+n}
\end{array}
\right]
\in \mathbb{R}^{n \times 9}
\end{equation}

where each row vector $\mathbf{p}_{t+k}$ contains the requested geographic coordinates and time information,

\begin{equation}
\mathbf{p}_{t+k} =
\big[
lat,\, lon,\, alt,\,
\sin(doy),\, \cos(doy),\,
\sin(ut),\, \cos(ut),\,
\sin(lst),\, \cos(lst)
\big]_{t+k}.
\end{equation}
Here $lat$, $lon$, and $alt$ denote latitude, longitude, and altitude, respectively, while 
$doy$, $ut$, and $lst$ denote day-of-year, universal time, and local solar time. 
The sine and cosine representations ensure temporal continuity.

\subsubsection{Historical information}
The historical input sequence is defined as
\begin{equation}
\mathbf{X}^{(h)}_t =
\left[
\begin{array}{c}
\mathbf{x}_{t-m+1}\\
\mathbf{x}_{t-m+2}\\
\vdots\\
\mathbf{x}_{t}
\end{array}
\right]
\in \mathbb{R}^{m \times (9+2n)}
\end{equation}

where the $2n$ terms correspond to JB2008 and NRLMSISE-00 baselines evaluated at $n$ requested locations, and the remaining 9 terms correspond to space-weather drivers.

Each historical feature vector $\mathbf{x}_k$ consists of empirical model density estimates and space weather indices:
\begin{equation}
\mathbf{x}_k =
\big[
\log_{10}\rho_{JB}(t_k,\mathbf{p}_{t+1:t+n}),\;
\log_{10}\rho_{MSIS}(t_k,\mathbf{p}_{t+1:t+n}),\;
\mathbf{s}(t_k),\;
\mathbf{g}(t_k),\;
\mathbf{w}(t_k)
\big],
\end{equation}
where $\rho_{JB}$ and $\rho_{MSIS}$ are density estimates from the empirical models JB2008 and NRLMSISE-00, evaluated at the requested future locations using solar and geomagnetic inputs at time $t_k$. 
The vectors $\mathbf{s}(t_k)$, $\mathbf{g}(t_k)$, and $\mathbf{w}(t_k)$ collect the space weather driving parameters:
\begin{align}
\mathbf{s}(t_k) &= [F_{10.7},\, F_{10.7A},\, F_{30}]_{t_k}\\
\mathbf{g}(t_k) &= [Dst,\, Ap30]_{t_k}\\
\mathbf{w}(t_k) &= [B_z,\, v,\, \rho_{proton},\, AE]_{t_k}
\end{align}
These parameters collectively represent solar irradiance variability, geomagnetic energy input, and solar wind parameters, which are the dominant external drivers of thermospheric density variability. Incorporating them enables the model to account for both long-term solar conditions and rapid storm-time disturbances.

\subsubsection{Dataset dimensions}
For a training dataset containing $T$ samples, the historical and future input tensors have dimensions
\begin{equation}
\mathbf{X}^{(h)} \in \mathbb{R}^{T \times m \times (9+2n)}, \qquad
\mathbf{X}^{(f)} \in \mathbb{R}^{T \times n \times 9}
\end{equation}
and the training target is
\begin{equation}
\mathbf{Y} \in \mathbb{R}^{T \times n}
\end{equation}
This input design enables the model to utilize empirical density baselines in conjunction with space weather indices, while conditioning forecasts on the specified future times and locations.

\subsubsection{Data normalization}

All input features and target variables are normalized using standardization (zero mean and unit variance). For each variable $x$, the normalized value $\tilde{x}$ is computed as

\begin{equation}
\tilde{x} = \frac{x - \mu}{\sigma}
\end{equation}

where $\mu$ and $\sigma$ denote the mean and standard deviation, respectively. The normalization statistics are computed separately for each input feature and the target variable using only the training dataset to avoid data leakage, and the same transformation is applied to the validation and test datasets. During evaluation, the predicted outputs are transformed back to the original physical scale using the inverse transformation for interpretation and performance assessment.

\subsection{Neural Network Architecture}

\begin{figure}[!t]
    \centering
    \includegraphics[width=1\linewidth]{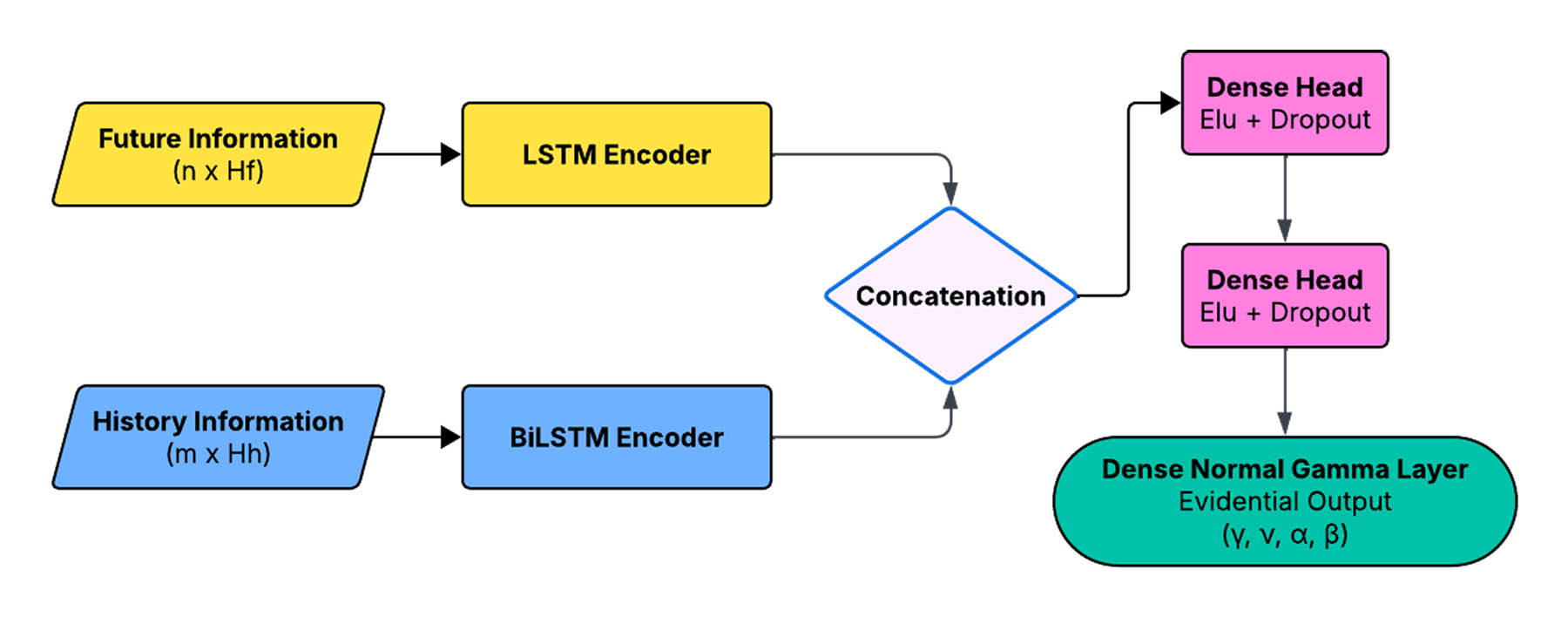}
    \caption{Neural network architecture of the AETHER-P$^3$}
    \label{fig:NN}
\end{figure}

Figure~\ref{fig:NN} illustrates the architecture of the proposed evidential thermospheric density forecasting model. The network accepts two inputs: a historical information tensor $\mathbf{X}^{(h)}_t \in \mathbb{R}^{m \times H_h}$ and a future requested time-location tensor $\mathbf{X}^{(f)}_t \in \mathbb{R}^{n \times H_f}$. The historical branch is encoded using a bidirectional long short-term memory (BiLSTM) layer with 192 hidden units per direction and a dropout rate of 0.20. In parallel, the future branch is encoded using a unidirectional LSTM layer with 128 hidden units and a dropout rate of 0.20. These encoders summarize the temporal evolution of prior thermospheric and space weather conditions as well as the requested forecast context.

The two context vectors are concatenated and batch-normalized to stabilize feature scales before being passed to a fully connected prediction head. The head consists of two dense layers, each with 256 and 192 neurons, respectively, utilizing ELU activation functions and dropout rates of 0.15. An additional dropout layer with a rate of 0.10 is applied before the output layer. A mild $\ell_2$ regularization with coefficient $1\times10^{-4}$ is applied to the dense layer kernels to reduce overfitting.

To jointly predict the multi-step density sequence and its associated uncertainty, the final layer is a Normal-Gamma evidential regression head. This head outputs the distribution parameters for each forecast step, enabling both density predictions and estimates of predictive uncertainty. The key architecture and training hyperparameters of the proposed model are summarized in Table~\ref{tab:Hyperparameters}.

\begin{table}[ht!]
\centering
\begin{tabular}{l c}
\hline
\textbf{Hyperparameter} & \textbf{Setting} \\
\hline
BiLSTM hidden units per direction & 192 \\
LSTM hidden units & 128 \\
LSTM dropout & 0.20 \\
Dense layer sizes & 256, 192 \\
Dense dropout & 0.15, 0.15 \\
Head dropout & 0.10 \\
Dense $\ell_2$ regularization coefficient & $1\times10^{-4}$ \\
Activation function & ELU \\
Batch size & 256 \\
Learning rate & $5\times10^{-5}$ \\
\hline
\end{tabular}
\caption{Key hyperparameters of the neural network model}
\label{tab:Hyperparameters}
\end{table}

\subsection{Data}

\begin{figure}[!t]
    \centering
    \includegraphics[width=\linewidth]{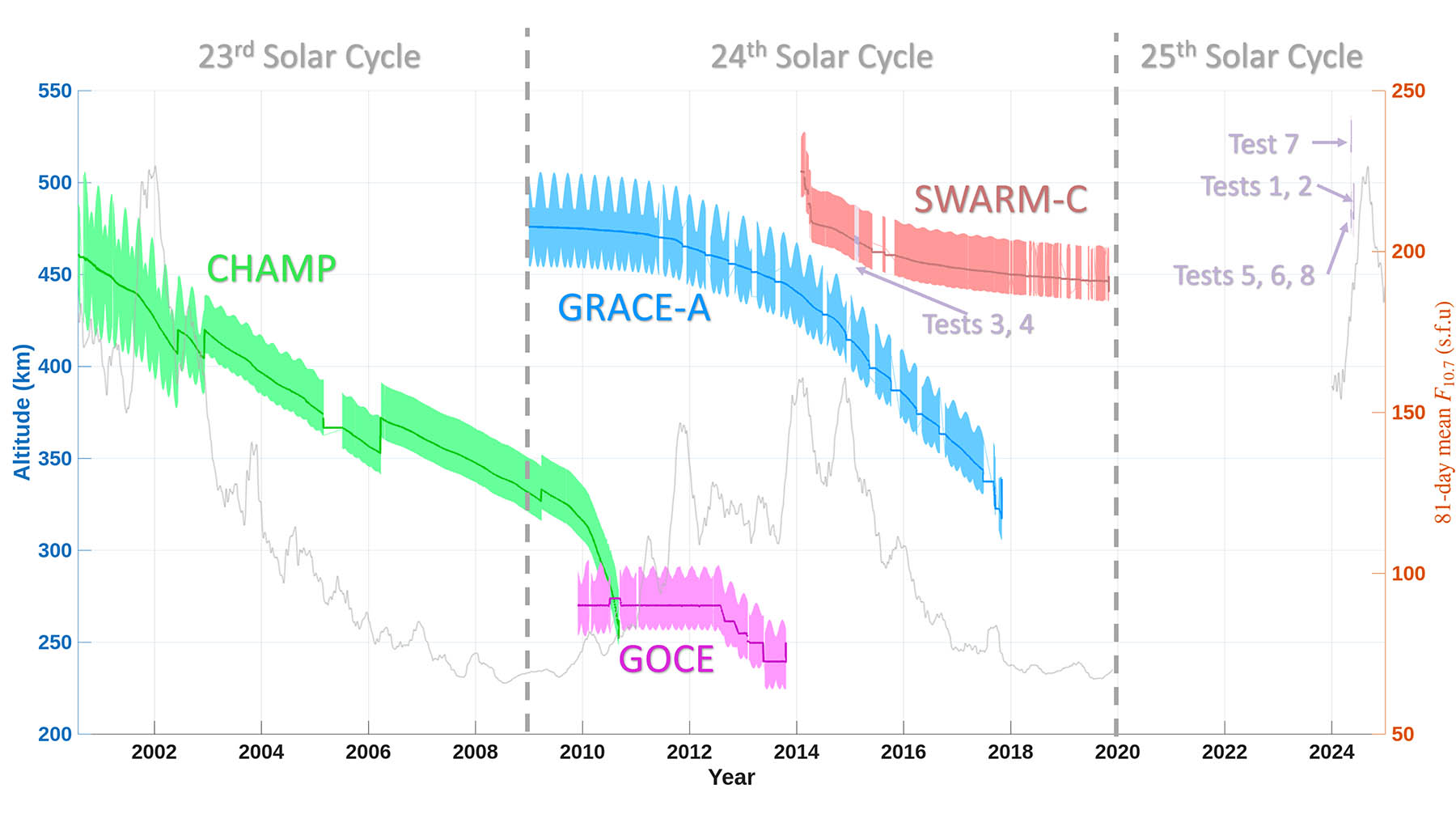}
    \caption{Satellite altitude evolution and corresponding $F_{10.7A}$ index for all training and test datasets.}
    \label{fig:satellite_alt}
\end{figure}

For the global forecasting model, thermospheric density measurements derived from onboard accelerometers are treated as ground truth. Figure~\ref{fig:satellite_alt} shows the altitude evolution of the training and test satellites together with the corresponding $F_{10.7A}$ index, highlighting the coverage of different solar activity conditions across solar cycles 23–25. As summarized in Table~\ref{tab:train_period}, datasets from the CHAMP, GRACE-A, GOCE, and SWARM-C satellites, spanning solar cycles 23 and 24, are used for model training. All datasets are obtained from the European Space Agency \cite{ESA, GOCE}. Test intervals overlapping the nominal training periods were excluded from the training dataset before sample construction. For each independent test case, all samples whose historical input window or forecast target window overlapped with the corresponding test interval were removed. This exclusion accounts for the 3-hour input history and 6-hour forecast horizon and prevents temporal leakage between training and testing samples. In particular, although the nominal SWARM-C training period spans 2014/02/01–2019/12/31, the SWARM-C Test 4 interval, 2015/02/01–2015/02/28, and all samples with overlapping input or forecast windows were excluded from training. The validation set was then constructed from the remaining training data after these exclusions, ensuring no temporal overlap with the independent test cases.

\begin{table}[!ht]
\centering
\caption{Training dataset periods.}
\label{tab:train_period}
\begin{tabular}{lc}
\hline
\textbf{Satellite} & \textbf{Data period} \\
\hline
CHAMP   & 2000/07/29 -- 2010/09/04 \\
GRACE-A & 2009/01/01 -- 2017/10/31 \\
GOCE    & 2009/11/01 -- 2013/10/20 \\
SWARM-C & 2014/02/01 -- 2019/12/31 \\
\hline
\end{tabular}
\end{table}

The data resolution is 10 minutes. The input horizon is 3 hours (18 time steps) and the forecasting horizon is 6 hours (36 time steps). Test datasets are constructed from independent observations from GRACE-FO and SWARM-A/B/C. 

During training, sequences containing abnormal density values (e.g., $>10^{33}\,\mathrm{kg\,m^{-3}}$) are removed. During testing, isolated abnormal measurements are excluded from metric computation, while the remaining valid samples in the sequence are retained to obtain more available test data samples.

\begin{table}[!ht]
\centering
\caption{Definition of test cases used for model evaluation.}
\label{tab:test_definition}
\begin{tabular}{l l c l c l}
\hline
\textbf{Test} & \textbf{Satellite} & \textbf{Period} & \textbf{Category} & \textbf{Min Dst (nT)} & \textbf{Comparison models} \\
\hline
Test 1 & SWARM-A & 2024/05/24--31 & Quiet      & $-27$  & BGMA, WAM-IPE \\
Test 2 & SWARM-C & 2024/05/24--31 & Quiet      & $-27$  & BGMA, WAM-IPE \\
Test 3 & SWARM-A & 2015/02/01--28 & Moderate   & $-69$  & C/DA NRLMSISE-00 \\
Test 4 & SWARM-C & 2015/02/01--28 & Moderate   & $-69$  & C/DA NRLMSISE-00 \\
Test 5 & GRACE-FO & 2024/05/10--13 & Extreme    & $-406$ & WAM-IPE \\
Test 6 & SWARM-A & 2024/05/10--13 & Extreme    & $-406$ & WAM-IPE \\
Test 7 & SWARM-B & 2024/05/10--13 & Extreme    & $-406$ & WAM-IPE \\
Test 8 & SWARM-C & 2024/05/10--13 & Extreme    & $-406$ & WAM-IPE \\
\hline
\end{tabular}
\end{table}

Table~\ref{tab:test_definition} summarizes the test cases. Tests~1–2 correspond to geomagnetically quiet periods and are compared against BGMA and WAM-IPE \cite{pan2025interpretable, zhan2024quantifying}. Tests~3–4 follow the benchmark test configuration used in the C/DA NRLMSISE-00 study, representing moderate geomagnetic activity \cite{forootan2022forecasting}. Tests~5–8 cover the May~2024 Gannon extreme storm event and are used to assess model performance under severe space weather conditions \cite{parker2024satellite}.

\subsection{Evaluation Metrics}

Model performance is evaluated by forecasting accuracy and the reliability of uncertainty estimation. The metrics are computed by aggregating all forecast steps within the 6-hour horizon over the full evaluation interval for each test case.

\subsubsection{Forecast accuracy}

Forecast accuracy is evaluated using the Pearson correlation coefficient ($R$), the root-mean-square error (RMSE), and the relative error (RE). 
Let $\hat{\rho}_i$ and $\rho_i$ denote the predicted and true thermospheric densities at sample $i$, and let $N$ be the total number of evaluation samples. 
The metrics are defined as:

\begin{equation}
R = 
\frac{\sum_{i=1}^{N} (\rho_i - \bar{\rho})(\hat{\rho}_i - \bar{\hat{\rho}})}
{\sqrt{\sum_{i=1}^{N} (\rho_i - \bar{\rho})^2}
 \sqrt{\sum_{i=1}^{N} (\hat{\rho}_i - \bar{\hat{\rho}})^2}},
\end{equation}

\begin{equation}
RMSE = \sqrt{\frac{1}{N} \sum_{i=1}^{N} (\hat{\rho}_i - \rho_i)^2},
\end{equation}

\begin{equation}
RE = \frac{1}{N} \sum_{i=1}^{N} 
\left| \frac{\hat{\rho}_i - \rho_i}{\rho_i} \right|.
\end{equation}

Here $\bar{\rho}$ and $\bar{\hat{\rho}}$ are the sample means of the true and predicted densities. 
An ideal forecast yields $R$ close to unity and RMSE and RE approaching zero.

\subsubsection{Uncertainty evaluation}

The predictive uncertainty quality is assessed using the coverage rate of the uncertainty interval $2\sigma$ and the mean absolute calibration error (MACE). 
Let $\mu_i$ and $\sigma_i$ denote the mean and standard deviation predicted at sample $i$. The $2\sigma$ coverage rate is defined as

\begin{equation}
CR_{2\sigma} = 
\frac{1}{N} \sum_{i=1}^{N} 
\mathbb{I}\!\left( \rho_i \in [\mu_i - 2\sigma_i,\, \mu_i + 2\sigma_i] \right) \times 100\%,
\end{equation}

where $\mathbb{I}(\cdot)$ is the indicator function. For well-calibrated Gaussian uncertainties, $\mathrm{CR}_{2\sigma}$ should be close to $95\%$.

Calibration across multiple confidence levels is further quantified by the mean absolute calibration error,

\begin{equation}
MACE = \frac{1}{n_C} \sum_{k=1}^{n_C} 
\left| C_k - P_k \right|,
\end{equation}

where $C_k$ denotes the nominal confidence level (taken as $[5\%,10\%,\ldots,95\%,96\%,99\%]$), and $P_k$ is the empirical fraction of samples whose true values fall within the corresponding predictive credible interval. A lower MACE indicates better calibration of uncertainty.

\section{Results and Discussion}

This section presents a comprehensive evaluation of the proposed AETHER-P$^3$ forecasting framework using independent satellite test cases spanning geomagnetically quiet, moderate, and extreme storm conditions. Model performance is assessed in terms of both deterministic forecast accuracy and probabilistic uncertainty reliability, reflecting the dual objectives of accurate density prediction and operationally meaningful uncertainty quantification.

\subsection{General model performance}
To mitigate the impact of stochastic variability in neural network training, ten independent model instances were trained using different random seeds. For each test case, the predictive outputs from all trained models were first averaged to form an ensemble-mean forecast at every prediction horizon. Subsequently, forecast accuracy and uncertainty metrics were computed using these ensemble-mean predictions. This evaluation strategy reflects the effective performance of an ensemble evidential forecasting system and avoids bias that may arise from directly averaging metric values across individual model realizations.

\begin{table}[!ht]
\centering
\caption{Overall performance of the AETHER-P$^3$ model across all test cases.}
\label{tab:general_performance}
\begin{tabular}{l l c c c c c}
\hline
\textbf{Test} & \textbf{Condition} & \textbf{R} & \textbf{RMSE ($kg/m^3$)} & \textbf{RE} & \textbf{CR$_{2\sigma}$} & \textbf{MACE} \\
\hline
Test 1 & Quiet    & 0.9556 & $2.54\times10^{-13}$ & 0.1919 & 99.32 & 0.0528 \\
Test 2 & Quiet    & 0.9581 & $2.47\times10^{-13}$ & 0.1979 & 99.53 & 0.0579 \\
Test 3 & Moderate & 0.9306 & $1.85\times10^{-13}$ & 0.1311 & 99.75 & 0.1027 \\
Test 4 & Moderate & 0.9353 & $1.83\times10^{-13}$ & 0.1284 & 99.68 & 0.1060 \\
Test 5 & Extreme  & 0.8860 & $5.96\times10^{-13}$ & 0.3504 & 95.05 & 0.0106 \\
Test 6 & Extreme  & 0.9001 & $5.94\times10^{-13}$ & 0.2780 & 98.30 & 0.0547 \\
Test 7 & Extreme  & 0.8956 & $4.11\times10^{-13}$ & 0.2269 & 94.96 & 0.0484 \\
Test 8 & Extreme  & 0.9018 & $5.91\times10^{-13}$ & 0.2807 & 98.54 & 0.0510 \\
\hline
\end{tabular}
\end{table}

Table~\ref{tab:general_performance} summarizes the overall seed-averaged performance of the proposed forecasting model across all test cases. The results demonstrate consistently strong deterministic forecasting skill across geomagnetically quiet, moderate, and extreme conditions, while maintaining reliable uncertainty quantification. Under quiet periods (Tests~1--2), the model achieves high correlation coefficients ($R>0.95$) and low RMSE values on the order of $2.5\times10^{-13} kg/m^3$, indicating excellent agreement between predicted and observed thermospheric density. During moderate geomagnetic activity (Tests~3--4), forecast skill decreases modestly ($R\approx0.93$), yet relative errors remain below 14\%, demonstrating stable performance under increased thermospheric variability. As expected, deterministic accuracy degrades further during extreme storm conditions (Tests~5--8), with correlation coefficients in the range $R=0.89$--$0.90$ and increased RMSE; however, predictive skill remains robust given the highly nonlinear and rapidly evolving storm-time dynamics.

\begin{figure}[!t]
    \centering
    \includegraphics[width=1\linewidth]{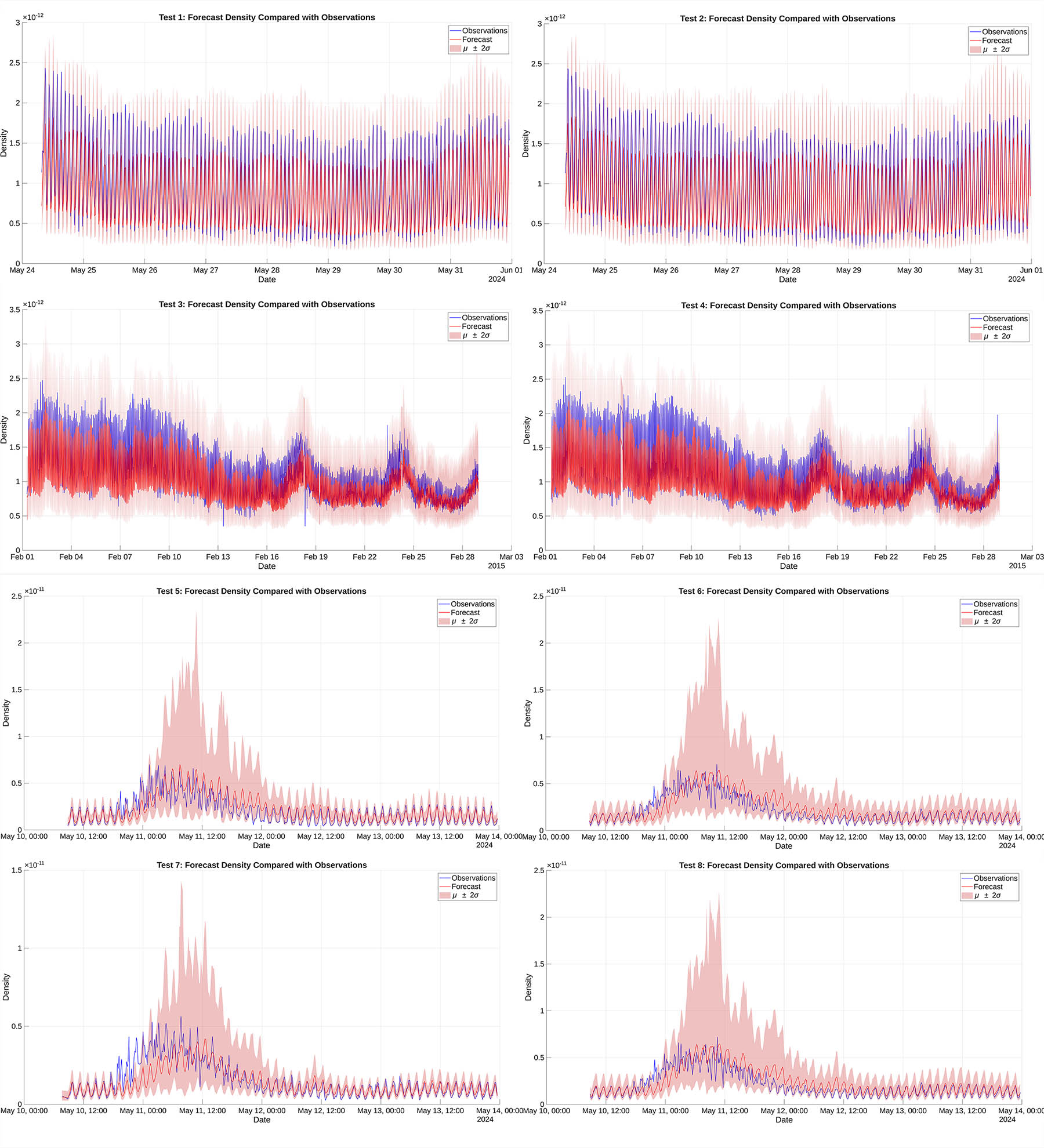}
    \caption{Time series comparison of ensemble-mean forecast and on-orbit thermospheric density observations along the satellite orbit for all test cases, with shaded regions indicating the predicted $\mu \pm 2\sigma$ uncertainty intervals.}
    \label{fig:density_all}
\end{figure}

Beyond aggregate statistics, Figure~\ref{fig:density_all} provides representative time-series comparisons between the ensemble-mean forecasts and accelerometer-derived observations for all test cases, together with the associated $\mu \pm 2\sigma$ predictive uncertainty intervals. The observations shown in these plots are on-orbit thermospheric densities derived from satellite-borne accelerometer measurements. Under geomagnetically quiet periods (Tests~1--2), the model closely tracks observed density variations along the satellite trajectories, accurately reproducing both diurnal oscillations and longer-timescale background trends. The predicted uncertainty intervals remain relatively narrow and consistently enclose the observations. This reflects stable external forcing and low intrinsic variability.

For moderate geomagnetic activity (Tests~3--4), increased short-term variability and intermittent density enhancements are evident. While forecast errors increase modestly, the model continues to capture the dominant temporal structure of the density evolution. Importantly, the predicted uncertainty intervals broaden during periods of enhanced variability, indicating that the evidential framework adapts uncertainty magnitude in response to changing geophysical conditions rather than maintaining a fixed confidence envelope.

During extreme storm conditions associated with the May~2024 Gannon event (Tests~5--8), thermospheric density exhibits rapid and highly nonlinear enhancements. Although deterministic forecast accuracy degrades during the peak storm phase, the predicted uncertainty intervals expand substantially and continue to encompass the majority of observed density excursions. This behavior is physically consistent with increased epistemic and aleatoric uncertainty under severe geomagnetic forcing and demonstrates that the proposed framework provides meaningful uncertainty information even when point prediction errors increase.
   
\begin{table}[!ht]
        \centering
        \caption{Performance in Terms of Pearson Correlation Coefficient (R) for Tests 5–8}
        \begin{tabular}{lccc}
        \hline
        \textbf{Test} & \textbf{Initial Phase} & \textbf{Main Phase} & \textbf{Recovery Phase} \\
        \hline
        Test 5 & 0.9791 & 0.7901 & 0.9706 \\
        Test 6 & 0.9325 & 0.8034 & 0.9413 \\
        Test 7 & 0.9171 & 0.7726 & 0.8989 \\
        Test 8 & 0.9719 & 0.8028 & 0.9405 \\     
        \hline
        \end{tabular}
        \label{tab:test5_8_results}
\end{table}

Table~\ref{tab:test5_8_results} summarizes the Pearson correlation coefficient (R) for each phase across Tests 5–8. The initial phase is defined as the period before 15:00 May 10, 2024; the main phase spans from 15:00 May 10 to 00:00 May 13, 2024; and the recovery phase covers the subsequent period.

During the initial phase, the model achieves consistently high correlation (R $\approx$ 0.92--0.98), indicating that it accurately captures the onset of thermospheric density enhancement associated with increasing geomagnetic activity. This suggests that the model effectively leverages upstream solar wind inputs together with geomagnetic activity information to capture early-stage forcing. During the main phase, the correlation decreases (R $\approx$ 0.77--0.80), reflecting the increased complexity and nonlinearity of thermospheric dynamics under strong geomagnetic forcing. As shown in Tests 5–8 in Figure~4, the model captures the rapid density enhancement and the timing of the peak response with minimal temporal lag. However, the peak amplitudes are not always accurately reproduced, with instances of both underestimation and overestimation. Given that this phase corresponds to the most intense period of the storm, such deviations are expected due to the highly dynamic and nonlinear system response. During the recovery phase, the model performance improves again (R $\approx$ 0.90--0.97), indicating that it accurately tracks the gradual decay of thermospheric density as geomagnetic activity subsides. The model reproduces both the large-scale trend and periodic variability during this phase. In addition, the predictive uncertainty ($\mu \pm 2\sigma$) expands substantially during the main phase, indicating reduced confidence in the model predictions under extreme storm-time conditions, and contracts during the initial and recovery phases. This behavior is consistent with the increased complexity and reduced predictability of thermospheric dynamics during the storm main phase. Overall, these results demonstrate that the AETHER-P$^3$ model effectively captures the key features of storm-time thermospheric density evolution across all phases, including rapid response to geomagnetic forcing, accurate peak timing, and realistic recovery behavior.

\begin{figure}[!t]
    \centering
    \includegraphics[width=0.72\linewidth]{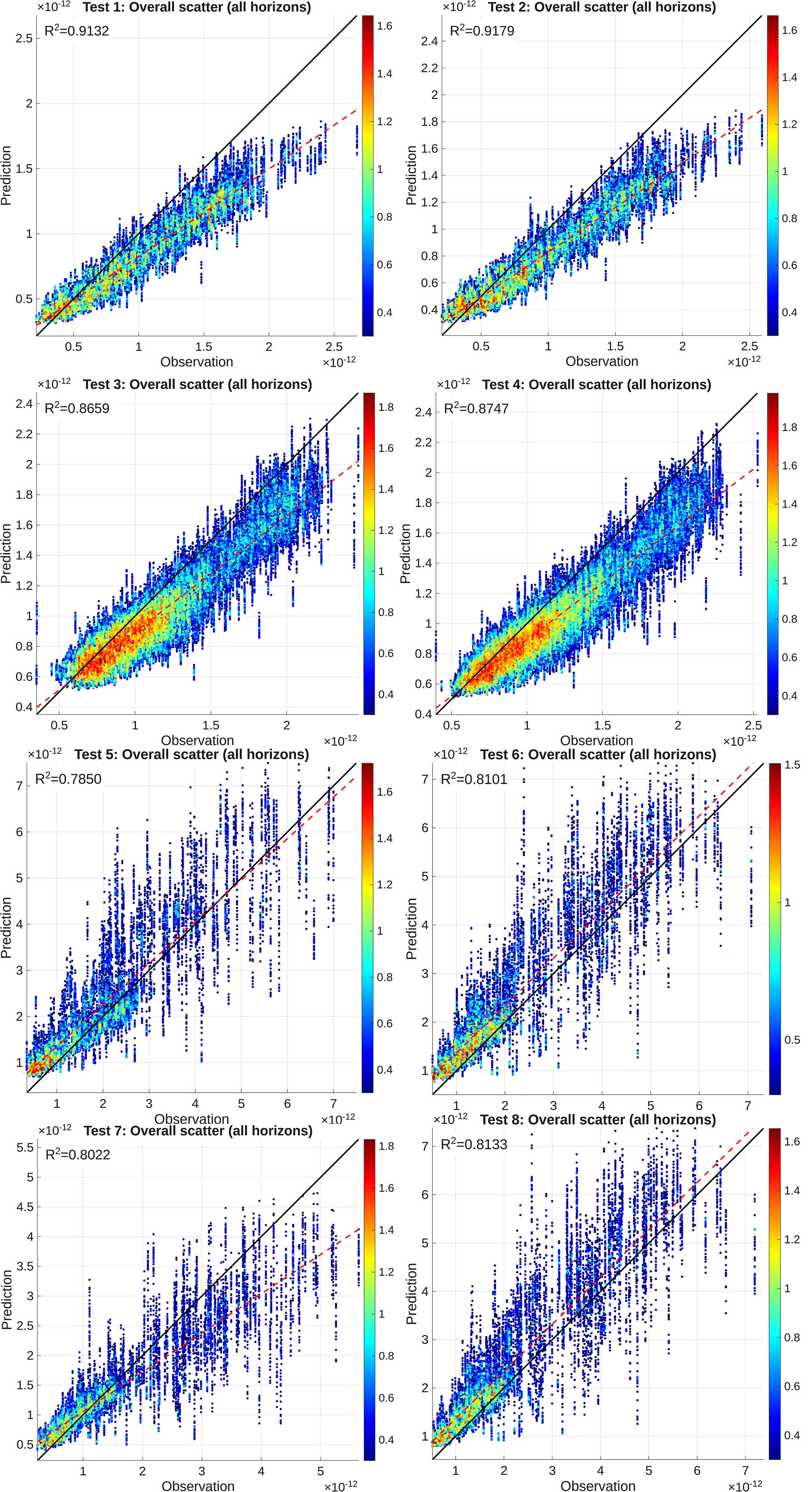}
    \caption{Scatter plots comparing predicted and observed thermospheric density for all test cases. The solid black line indicates the 1:1 reference line, and the dashed red line represents the first-order polynomial fit.}
    \label{fig:scatter_all}
\end{figure}

Deterministic prediction accuracy aggregated over all forecast horizons is further illustrated by the scatter plots in Figure~\ref{fig:scatter_all}, which compare predicted and observed thermospheric densities across all test cases. Due to the multi-step forecasting framework, multiple predictions can correspond to the same observation, as forecasts generated from different input windows and lead times may target the same time point. Under quiet periods, the scatter distributions closely follow the 1:1 reference line, indicating minimal systematic bias and strong consistency across the prediction range. Moderate activity cases exhibit slightly increased dispersion while maintaining tight clustering, consistent with the modest reduction in correlation observed in Table~\ref{tab:general_performance}. During extreme storm conditions, increased scatter and mild underestimation at higher density values are observed, reflecting the challenges of forecasting rapidly evolving storm-time responses.

\begin{figure}[!t]
    \centering
    \includegraphics[width=0.72\linewidth]{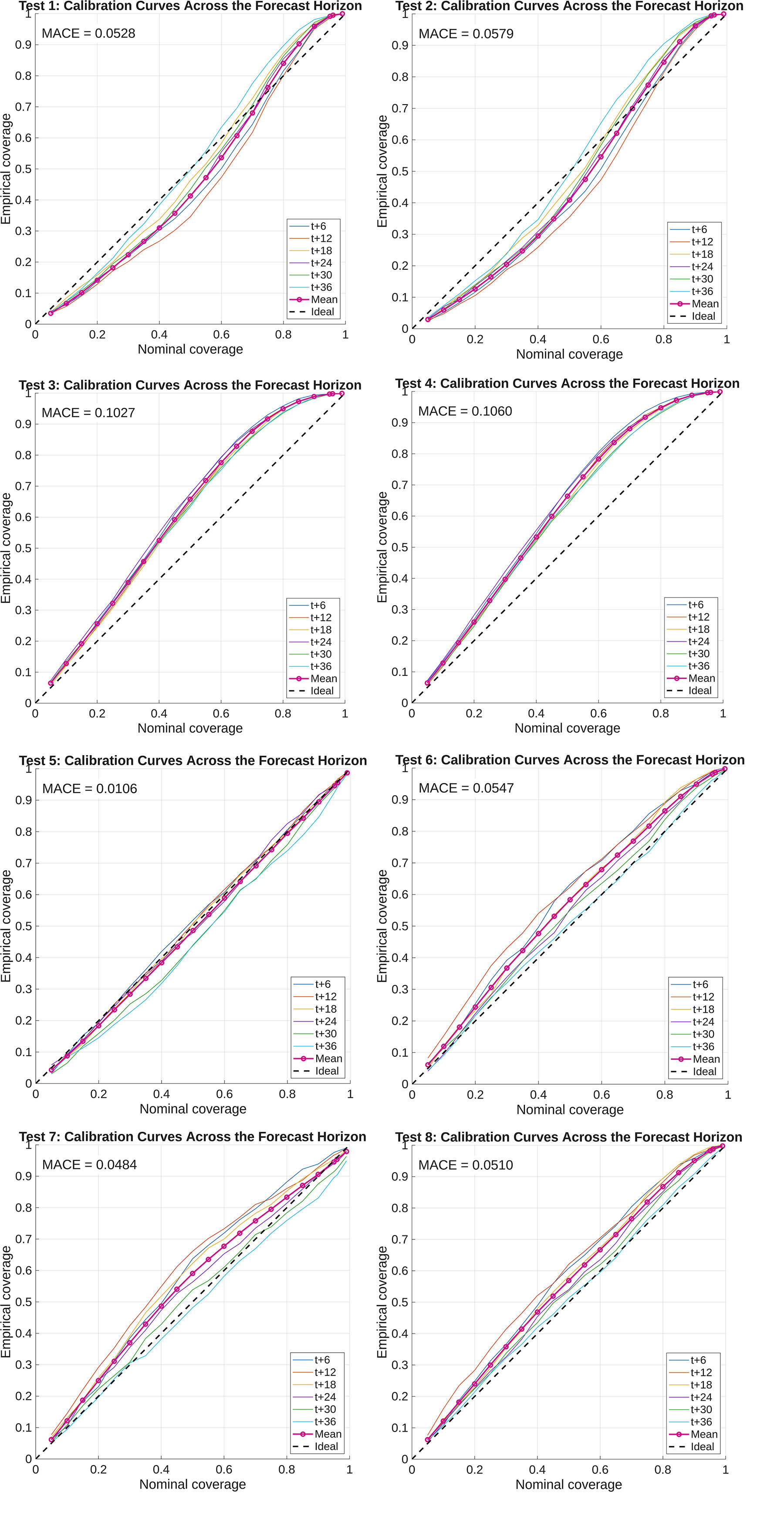}
    \caption{Calibration curve for all test cases.}
    \label{fig:calibration_all}
\end{figure}

The reliability of the predicted uncertainty is assessed using coverage calibration curves shown in Figure~\ref{fig:calibration_all}. Across all test cases, empirical coverage closely follows the ideal diagonal reference, indicating well-calibrated predictive uncertainty across multiple confidence levels. While moderate activity cases (Tests~3--4) exhibit slightly larger deviations from perfect calibration, particularly at intermediate nominal coverage levels, overall miscalibration remains limited, with MACE values on the order of 0.10. Notably, under extreme storm conditions (Tests~5--8), the model maintains robust uncertainty reliability, with $2\sigma$ coverage rates remaining close to or above the nominal 95\% level and consistently low MACE values (approximately 0.05). These results demonstrate that the evidential framework preserves statistically consistent uncertainty quantification even under severe space weather conditions.

Overall, the results in Table~\ref{tab:general_performance} and Figures~\ref{fig:density_all}--\ref{fig:calibration_all} demonstrate that AETHER-P$^3$ achieves strong deterministic accuracy and reliable uncertainty quantification across a broad range of geomagnetic conditions. The model exhibits consistently high correlation, low physical-domain errors, and near-nominal uncertainty coverage when performance is aggregated over all forecast horizons and test intervals. At the same time, these results should be interpreted within the altitude range represented by the independent validation cases. The present test cases primarily cover the altitude region where accelerometer-derived satellite data are most abundant and where the model is therefore expected to be most reliable, particularly the SWARM and GRACE/GRACE-FO orbital regimes.

To provide deeper insight into model behavior under different space weather regimes and to benchmark performance against existing approaches, the following subsections examine forecasting skill under geomagnetically quiet periods (Section~3.2), moderate geomagnetic activity (Section~3.3), and comparative performance against physics-based models under both quiet and storm-time conditions (Section~3.4).

\subsection{Performance under geomagnetically quiet periods}

\begin{table}[!ht]
\centering
\caption{Per-horizon correlation performance of the AETHER-P$^3$ model for Tests~1 and~2.}
\label{tab:horizon_R_R2_bgma}
\begin{tabular}{c cc cc c}
\hline
\textbf{Forecast Horizon} 
& \multicolumn{2}{c}{\textbf{Test 1 (SWARM-A)}} 
& \multicolumn{2}{c}{\textbf{Test 2 (SWARM-C)}} 
& \textbf{BGMA} \\
& $R$ & $R^2$ & $R$ & $R^2$ & $R^2$ \\
\hline
t+1  (10 min)  & 0.9490 & 0.9005 & 0.9503 & 0.9030 & 0.903 \\
t+2  (20 min)  & 0.9495 & 0.9016 & 0.9513 & 0.9050 & -- \\
t+3  (30 min)  & 0.9497 & 0.9018 & 0.9506 & 0.9036 & 0.850 \\
t+4  (40 min)  & 0.9514 & 0.9051 & 0.9533 & 0.9088 & -- \\
t+5  (50 min)  & 0.9567 & 0.9152 & 0.9580 & 0.9177 & -- \\
t+6  (1 hr)    & 0.9557 & 0.9134 & 0.9566 & 0.9151 & 0.805 \\
t+12 (2 hr)    & 0.9539 & 0.9100 & 0.9560 & 0.9139 & -- \\
t+18 (3 hr)    & 0.9571 & 0.9160 & 0.9594 & 0.9204 & -- \\
t+24 (4 hr)    & 0.9602 & 0.9219 & 0.9632 & 0.9277 & -- \\
t+30 (5 hr)    & 0.9598 & 0.9213 & 0.9630 & 0.9275 & -- \\
t+36 (6 hr)    & 0.9574 & 0.9166 & 0.9611 & 0.9237 & -- \\
\hline
\end{tabular}
\end{table}

Table~\ref{tab:horizon_R_R2_bgma} presents horizon-dependent correlation performance for the quiet-period test cases and includes BGMA benchmark results at matched forecast horizons. 
At the 10-minute forecasting horizon, the proposed model exhibits correlation performance comparable to the BGMA model. However, at longer horizons (30~min and 1~hr), the proposed model maintains substantially higher $R^2$ values, indicating superior temporal forecasting capability beyond short-term persistence. Furthermore, correlation coefficients of the proposed model remain above 0.95 across the entire 6-hour forecasting window for both Test~1 and Test~2. 
This sustained high-level performance suggests that the model effectively captures the slowly varying thermospheric response under geomagnetically quiet periods, where external solar and geomagnetic driving factors exhibit limited temporal variability. As a result, the global forecasting framework maintains stable and accurate multi-step predictions throughout the full test interval.

\subsection{Performance under moderate geomagnetic activity}

\begin{table}[t]
\centering
\small
\caption{Overall forecasting error comparison for Tests~3 and~4.}
\label{tab:mae_rmse_physical_test34}
\setlength{\tabcolsep}{6pt}
\renewcommand{\arraystretch}{1.15}

\begin{tabular}{l ccc ccc}
\hline
& \multicolumn{3}{c}{\textbf{Test 3 (SWARM-A)}}
& \multicolumn{3}{c}{\textbf{Test 4 (SWARM-C)}} \\
\textbf{Model} & \textbf{MAE} & \textbf{RMSE} & \textbf{RE}
& \textbf{MAE} & \textbf{RMSE} & \textbf{RE} \\

 & \textbf{($kg/m^3$)} & \textbf{($kg/m^3$)} & 
& \textbf{($kg/m^3$)} & \textbf{($kg/m^3$)} & \\
\hline

JB2008
& 1.554$\times 10^{-13}$ & 2.101$\times 10^{-13}$ & 0.1389
& 1.517$\times 10^{-13}$ & 2.033$\times 10^{-13}$ & 0.1359 \\

NRLMSISE-00
& 3.880$\times 10^{-13}$ & 4.452$\times 10^{-13}$ & 0.3391
& 4.038$\times 10^{-13}$ & 4.610$\times 10^{-13}$ & 0.3483 \\

AETHER-P$^3$
& 1.112$\times 10^{-13}$ & 1.465$\times 10^{-13}$ & 0.1052
& 1.100$\times 10^{-13}$ & 1.463$\times 10^{-13}$ & 0.1009 \\ \hline

\end{tabular}
\end{table}

Although empirical models do not provide true forecasts, their nowcasting performance could be used as a baseline for evaluating thermospheric density forecasting skill. Both NRLMSISE-00 and JB2008 are empirical nowcasting models that estimate thermospheric density using contemporaneous space-weather inputs available at the time of prediction. As such, they benefit from access to information that is not available to forecasting models and are therefore expected to exhibit higher accuracy. In this work, these empirical models are used as reference baselines rather than direct forecasting competitors. Following the evaluation strategy adopted in Forootan's forecasting studies, the performance of the proposed AETHER-P$^3$ model is assessed relative to empirical model nowcasts to provide a practical and consistent benchmark for forecasting capability \cite{forootan2022forecasting}.

Table~\ref{tab:mae_rmse_physical_test34} summarizes the overall forecasting performance in the physical density domain for the moderate-activity Tests~3 and~4 by using Mean Absolute Error (MAE), RMSE, and RE metrics. Results show that AETHER-P$^3$ achieves substantially lower errors than both baseline models, even though it is performing forecasting tasks, which is significantly more challenging than the nowcasting task performed by the empirical model. For Test~3 (SWARM-A), AETHER-P$^3$ attains an RMSE of \(1.465\times10^{-13} kg/m^3\), compared with \(4.452\times10^{-13} kg/m^3\) for NRLMSISE-00 and \(2.101\times10^{-13} kg/m^3\) for JB2008. For Test~4 (SWARM-C), the forecasting model yields an RMSE of \(1.463\times10^{-13} kg/m^3\), while NRLMSISE-00 and JB2008 produce RMSEs of \(4.610\times10^{-13} kg/m^3\) and \(2.033\times10^{-13} kg/m^3\), respectively, corresponding to RMSE reductions of approximately \(67\%\)–\(68\%\) relative to NRLMSISE-00 and \(28\%\)–\(30\%\) relative to JB2008. 

Furthermore, AETHER-P$^3$ exceeds the performance of the data-assimilative C/DA NRLMSISE-00 model, which reported RMSE reductions of \(53\%\) (SWARM-A) and \(56\%\) (SWARM-C) relative to NRLMSISE-00 under comparable benchmark conditions. These results demonstrate that AETHER-P$^3$ maintains high predictive accuracy during periods of moderate geomagnetic activity.

\subsection{Comparative performance against physics-based models}

\begin{table}[!t]
\centering
\small
\setlength{\tabcolsep}{6pt}
\renewcommand{\arraystretch}{1.15}
\caption{Performance comparison between WAM-IPE density estimates and the global forecasting model.}
\label{tab:wamipe}

\begin{tabular}{l l c c c}
\hline
\textbf{Test} & \textbf{Model} & \textbf{R} & \textbf{RMSE ($kg/m^3$)} & \textbf{RE} \\
\hline
Test 1 & WAM-IPE (Nearest)        & 0.9479 & 9.2577$\times 10^{-13}$ & 0.9916 \\
Test 1 & WAM-IPE (Interpolation)  & \textbf{0.9548} & 9.1274$\times 10^{-13}$ & 0.9883 \\
Test 1 & AETHER-P$^3$ 
       & 0.9490 & \textbf{2.7432$\times 10^{-13}$} & \textbf{0.2021} \\

Test 2 & WAM-IPE (Nearest)        & 0.9475 & 9.2213$\times 10^{-13}$ & 1.0583 \\
Test 2 & WAM-IPE (Interpolation)  & \textbf{0.9555} & 9.0760$\times 10^{-13}$ & 1.0530 \\
Test 2 & AETHER-P$^3$
       & 0.9503 & \textbf{2.6616$\times 10^{-13}$} & \textbf{0.2103} \\

Test 5 & WAM-IPE (Nearest)        & 0.6810 & 2.1306$\times 10^{-12}$ & 1.1416 \\
Test 5 & WAM-IPE (Interpolation)  & 0.6804 & 2.1093$\times 10^{-12}$ & 1.1374 \\
Test 5 & AETHER-P$^3$ 
       & \textbf{0.8868} & \textbf{7.2636$\times 10^{-13}$} & \textbf{0.3673} \\

Test 6 & WAM-IPE (Nearest)        & 0.7180 & 1.8026$\times 10^{-12}$ & 0.8713 \\
Test 6 & WAM-IPE (Interpolation)  & 0.7186 & 1.7778$\times 10^{-12}$ & 0.8635 \\
Test 6 & AETHER-P$^3$ 
       & \textbf{0.8993} & \textbf{7.8059$\times 10^{-13}$} & \textbf{0.2797} \\

Test 7 & WAM-IPE (Nearest)        & 0.7358 & 1.6434$\times 10^{-12}$ & 1.1004 \\
Test 7 & WAM-IPE (Interpolation)  & 0.7414 & 1.6478$\times 10^{-12}$ & 1.1078 \\
Test 7 & AETHER-P$^3$
       & \textbf{0.9136} & \textbf{5.0925$\times 10^{-13}$} & \textbf{0.2211} \\

Test 8 & WAM-IPE (Nearest)        & 0.7169 & 1.8091$\times 10^{-12}$ & 0.8878 \\
Test 8 & WAM-IPE (Interpolation)  & 0.7155 & 1.7901$\times 10^{-12}$ & 0.8803 \\
Test 8 & AETHER-P$^3$
       & \textbf{0.9012} & \textbf{7.7819$\times 10^{-13}$} & \textbf{0.2889} \\
\hline
\end{tabular}
\end{table}

The performance of AETHER-P$^3$ is benchmarked against the physics-based WAM-IPE model under both geomagnetically quiet periods (Tests~1--2) and extreme storm-time forcing (Tests~5--8). Because the testing period for Tests~3-4 is one month, which is much longer than for other tests, the computation of the forecast results of the WAM-IPE model is more intensive. Therefore, the results of Tests~3-4 will not be compared with the WAM-IPE model. Table~\ref{tab:wamipe} summarizes quantitative performance metrics for Tests~1--2 (geomagnetically quiet periods) and Tests~5--8, which correspond to the May~2024 Gannon extreme storm event. For WAM-IPE, two spatial sampling strategies are considered: a nearest-grid approach, which assigns the density value at the closest model grid point to the satellite position, and a three-dimensional interpolation approach, which estimates density at the exact satellite location using surrounding grid cells.

\begin{figure}[!t]
    \centering
    \includegraphics[width=1\linewidth]{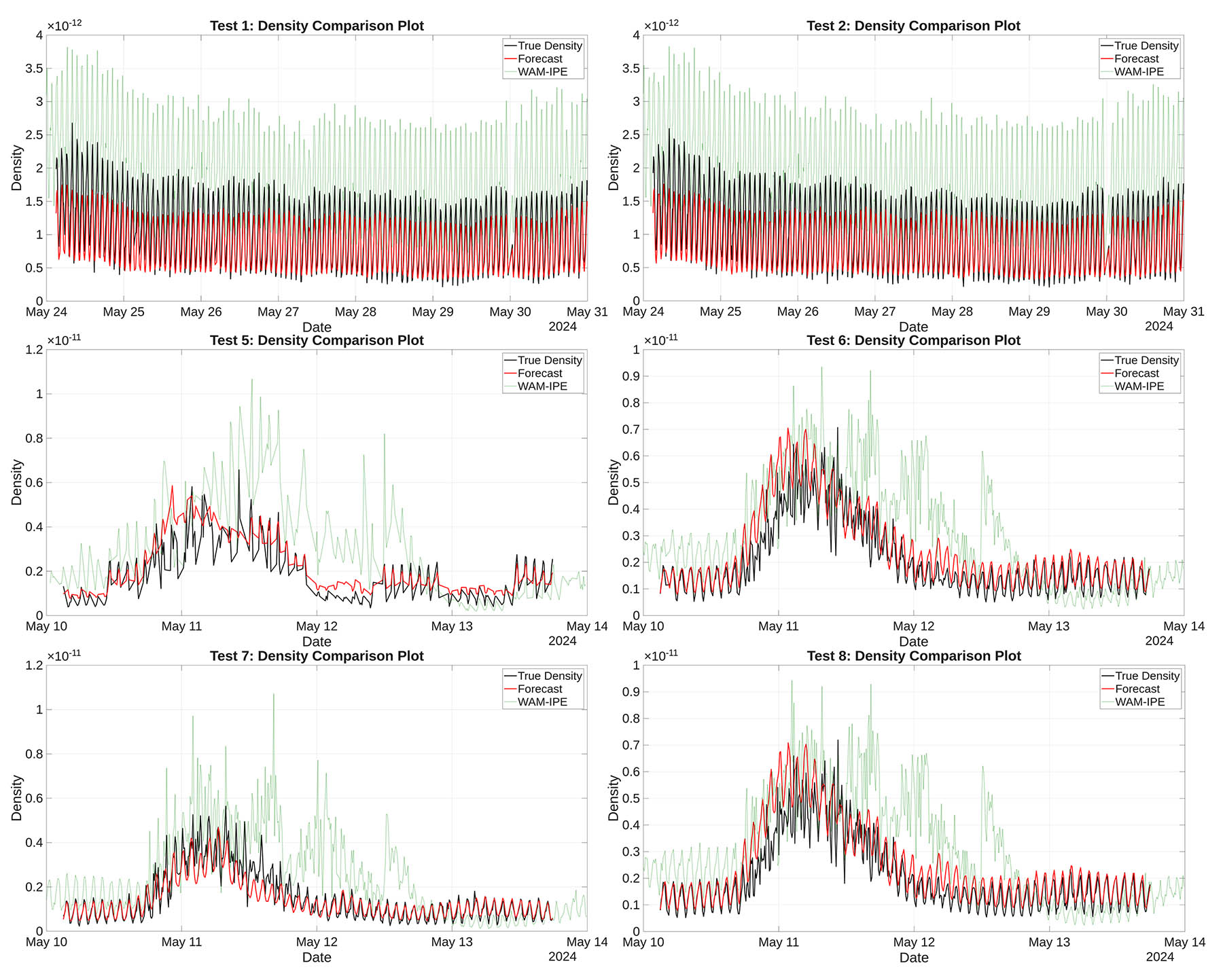}
    \caption{Time-series comparison of forecast, on-orbit thermospheric density observations derived from satellite-borne accelerometer measurements, and interpolated WAM-IPE density estimates for quiet and storm-time test cases.}
    \label{fig:wamipe_all}
\end{figure}

Consistent with prior studies and operational practice, the interpolated WAM-IPE density generally provides slightly improved agreement with satellite observations compared to the nearest-grid sampling, particularly in terms of correlation and magnitude error. Accordingly, the interpolated WAM-IPE results are adopted for the time-series comparisons shown in Figure~\ref{fig:wamipe_all}, as they represent a more accurate and physically consistent estimate of model density along the satellite trajectories.

Under geomagnetically quiet periods (Tests~1--2), the AETHER-P$^3$ model achieves correlation coefficients comparable to those of the interpolated WAM-IPE outputs ($R\approx0.95$), reflecting the relatively stable thermospheric response and slowly varying external forcing. Despite similar correlation performance, the forecasting model consistently yields substantially lower RMSE and relative error, reducing magnitude errors by approximately 70\% relative to WAM-IPE. This improvement indicates that the proposed data-driven framework more accurately captures absolute density levels along the satellite orbits, even when large-scale background conditions are well represented by physics-based models.

During extreme storm conditions (Tests~5--8), WAM-IPE performance degrades markedly, with correlation coefficients dropping to $R\approx0.68$--0.74 and RMSE values exceeding $1.6\times10^{-12}$. In contrast, the AETHER-P$^3$ model maintains robust deterministic skill, achieving correlations above 0.88 across all storm-time tests and reducing RMSE by approximately 60--70\% relative to WAM-IPE. Relative error is similarly reduced, highlighting improved tracking of rapid storm-time density enhancements.
 
Figure~\ref{fig:wamipe_all} provides time-series comparisons between the ensemble-mean forecasts, on-orbit thermospheric density observations derived from satellite-borne accelerometer measurements along the satellite orbit, and interpolated WAM-IPE density estimates for representative quiet and storm-time intervals. During the peak phase of the May~2024 storm, WAM-IPE exhibits delayed and attenuated responses to rapid density enhancements, while the proposed forecast model more closely follows the observed temporal evolution. These results demonstrate that, even relative to a state-of-the-art physics-based forecasting system, the proposed evidential deep learning framework delivers improved along-track density prediction during severe geomagnetic disturbances.

Overall, these results indicate that AETHER-P$^3$ generalizes effectively beyond climatological conditions and maintains strong predictive skill under highly nonlinear, rapidly evolving space weather forcing. This capability is particularly important for operational drag forecasting during geomagnetic storms, where timely and accurate density prediction along satellite trajectories is critical for risk-aware decision-making.

\section{Conclusions}

This study presents and evaluates AETHER-P$^3$, a machine-learning-based global thermospheric density forecasting framework designed for multi-step prediction with uncertainty quantification. AETHER-P$^3$ formulates thermospheric density forecasting as a conditional sequence-to-sequence regression problem, in which future density predictions are jointly conditioned on recent thermospheric and space-weather evolution and on a user-specified sequence of future times and locations. Historical inputs incorporate physically informed empirical density baselines from JB2008 and NRLMSISE-00 evaluated at the requested future locations, together with key solar, geomagnetic, and solar-wind drivers. A dual-branch recurrent neural network architecture, consisting of a BiLSTM encoder for historical context and an LSTM encoder for future request context, is coupled with a Normal-Gamma evidential regression head to provide both point predictions and predictive uncertainty. This improved forecasting performance can be interpreted from a physical perspective. Thermospheric density exhibits strong temporal memory due to the cumulative and delayed response to solar and geomagnetic forcing, governed by energy deposition, diffusion, and transport processes. By incorporating historical space weather drivers and upstream solar wind parameters, the proposed framework effectively captures the evolving thermospheric state, which is particularly beneficial under moderate geomagnetic conditions where the response remains structured and predictable.

Model performance was assessed using eight independent satellite test cases spanning geomagnetically quiet periods, moderate activity, and extreme storm-time forcing. To reduce stochastic variability associated with neural network training, all results were reported using ensemble-mean forecasts derived from ten independently trained model realizations. Under quiet periods (Tests~1--2), AETHER-P$^3$ achieved high deterministic forecasting skill, with correlation coefficients exceeding 0.95 and RMSE values on the order of $2.5\times10^{-13} kg/m^3$, while maintaining consistent performance across the full 6-hour forecast horizon. During moderate geomagnetic activity (Tests~3--4), the model sustained strong predictive skill ($R\approx0.93$). Although empirical models are limited to nowcasting, their predictive accuracy is expected to be higher than that of true forecasting models because they rely on contemporaneous and reliable space-weather inputs. By taking the same evaluation strategy as some previous forecasting studies, the AETHER-P$^3$ model will be compared with nowcasting empirical models to estimate its forecasting skill. Despite this advantage in reliable input data for the nowcasting model, the proposed AETHER-P$^3$ model exhibits substantially lower physical-domain errors than the empirical baseline models, outperforming JB2008 and NRLMSISE-00 in terms of MAE, RMSE, and relative error. These results demonstrate that AETHER-P$^3$ captures nontrivial temporal evolution beyond short-horizon persistence under slowly to moderately varying external forcing.

During the May~2024 extreme geomagnetic storm event (Tests~5--8), deterministic forecast skill degraded as expected due to the highly nonlinear and rapidly evolving thermospheric response. Nevertheless, AETHER-P$^3$ remained robust, achieving correlation coefficients in the range 0.89--0.90. In direct comparison with WAM-IPE density estimates sampled along satellite trajectories, AETHER-P$^3$ delivered markedly improved storm-time performance, reducing RMSE by approximately 60--70\%. These results indicate enhanced generalization under severe geomagnetic forcing and improved capability to track rapid storm-time density enhancements relative to state-of-the-art physics-based forecasting systems.

Uncertainty reliability was evaluated using $2\sigma$ coverage rate and mean absolute calibration error (MACE). Calibration curves across all test cases closely followed the ideal diagonal reference, indicating statistically consistent uncertainty quantification across multiple confidence levels. In particular, storm-time tests exhibited near-nominal $2\sigma$ coverage (approximately 95\%) and low MACE values (approximately 0.05), demonstrating that the evidential framework employed by AETHER-P$^3$ preserves reliable uncertainty estimates even when deterministic forecast errors increase. This reliability is critical for operational applications in which decision-making depends not only on expected thermospheric density but also on quantified forecast confidence.

Overall, AETHER-P$^3$ provides a practical, low-latency, and uncertainty-aware global thermospheric density forecasting capability that performs robustly across quiet-to-extreme geomagnetic regimes. By jointly improving multi-step forecasting accuracy and uncertainty reliability, the proposed framework enhances thermospheric density prediction and directly supports operational satellite drag prediction, conjunction assessment, and risk-informed space weather decision-making. This operational relevance should be interpreted within the altitude range supported by the available accelerometer-derived training data and independent validation cases. In its present form, AETHER-P$^3$ is practically validated for LEO density forecasting over approximately 300–520 km, with greatest confidence in the data-rich 400–520 km altitude range. Broader altitude applicability will require additional satellite observations and independent validation cases outside this range. Future work will focus on extending the forecast horizon, improving the representation of storm-time dynamics by incorporating forecasted space-weather drivers, expanding validation across broader altitude and solar-cycle conditions, and testing AETHER-P$^3$ using additional independent satellite data.

\section*{Data Availability Statement}

All data used in this study are obtained from publicly available sources. All the satellite datasets are obtained from the European Space Agency \cite{ESA, GOCE}.
The empirical thermospheric density models JB2008 and NRLMSISE-00 are accessed through open-source MATLAB implementations \cite{Mahooti2026-JBAtmosDensityModel, Mahooti2026-nrlmsise00}. 
Solar activity indices, including $F_{10.7}$ and $F_{10.7A}$, are obtained from \cite{kelso}, while the $F_{30}$ radio flux is provided by \cite{CLS_RadioFlux}. 
Geomagnetic activity indices include the $Dst$ index from \cite{chok} and the Ap30 index from \cite{Matzka2024Hpo}. 
Solar wind parameters, including $B_z$, solar wind speed $v$, proton density $\rho_{\mathrm{proton}}$, and the AE index, are obtained from the OMNI database \cite{OMNIWeb_Minute}. The programming scripts used to generate the results shown in this article are publicly archived on Zenodo and are available at \url{https://doi.org/10.5281/zenodo.20412490} \cite{wang2026software}.

\section*{Conflict of Interest declaration}
The authors declare that there are no conflicts of interest for this manuscript.

\acknowledgments
This research has been supported by the National Science Foundation, United States,
696 under Award 2149747, and NASA, United States, under Award 80NSSC24K0843.

\bibliography{reference}

\end{document}